\documentclass[twocolumn]{aastex631}

\usepackage{threeparttable}
\usepackage{here}

\shorttitle{The microvariability and wavelength dependence of polarization degree/angle of BL Lac}
\shortauthors{Imazawa et al.}

\begin{document}

\title{The microvariability and wavelength dependence of polarization degree/angle of BL Lacertae in the outburst 2020 to 2021}

\correspondingauthor{Ryo Imazawa}
\email{imazawa@astro.hiroshima-u.ac.jp}

\author[0000-0002-0643-7946]{Ryo Imazawa}
\affiliation{Department of Physics, Graduate School of Advanced Science and Engineering, Hiroshima University
Kagamiyama, 1-3-1 Higashi-Hiroshima, Hiroshima 739-8526, Japan}

\author[0000-0001-5946-9960]{Mahito Sasada}
\affiliation{Department of Physics, Tokyo Institute of Technology, 2-12-1 Ookayama, Meguro-ku, Tokyo 152-8551, Japan}
\affiliation{Hiroshima Astrophysical Science Center, Hiroshima University 1-3-1 Kagamiyama, Higashi-Hiroshima, Hiroshima 739-8526, Japan}

\author[0000-0001-9242-7973]{Natsuko Hazama}
\affiliation{Department of Physics, Graduate School of Advanced Science and Engineering, Hiroshima University
Kagamiyama, 1-3-1 Higashi-Hiroshima, Hiroshima 739-8526, Japan}

\author[0000-0002-0921-8837]{Yasushi Fukazawa}
\affiliation{Department of Physical Science, Hiroshima University 1-3-1  Kagamiyama, Higashi-Hiroshima Hiroshima 739-8526, Japan}
\affiliation{Department of Physics, Graduate School of Advanced Science and Engineering, Hiroshima University
Kagamiyama, 1-3-1 Higashi-Hiroshima, Hiroshima 739-8526, Japan}

\author{Tatsuya Nakaoka}
\affiliation{Hiroshima Astrophysical Science Center, Hiroshima University 1-3-1 Kagamiyama, Higashi-Hiroshima, Hiroshima 739-8526, Japan}

\author[0000-0001-6156-238X]{Hiroshi Akitaya}
\affiliation{Planetary Exploration Research Center, Chiba Institute of Technology, 2-17-1 Tsudanuma, Narashino, Chiba 275-0016, Japan}

\author[0000-0001-6099-9539]{Koji S. Kawabata}
\affiliation{Hiroshima Astrophysical Science Center, Hiroshima University 1-3-1 Kagamiyama, Higashi-Hiroshima, Hiroshima 739-8526, Japan}

\author[0000-0002-4375-254X]{Thomas Bohn}
\affiliation{Hiroshima Astrophysical Science Center, Hiroshima University 1-3-1 Kagamiyama, Higashi-Hiroshima, Hiroshima 739-8526, Japan}

\author[0000-0002-3884-5637]{Anjasha Gangopadhyay}
\affiliation{Hiroshima Astrophysical Science Center, Hiroshima University 1-3-1 Kagamiyama, Higashi-Hiroshima, Hiroshima 739-8526, Japan}

\begin{abstract}

We have obtained simultaneous and continuous photo-polarization observations of the blazar BL Lacertae in optical and near-infrared (NIR) bands during a historical outburst from 2020 to 2021. 
In total, fourteen nights of observations were performed where ten observations show microvariability on timescales of a few minutes to several hours.
This suggests a compact emission region, and the timescales are difficult to explain by a one-zone shock-in-jet model.
Moreover, we found significant differences in the polarization degree (PD) and angle between optical and NIR bands. 
Nine nights showed a PD in the optical band that is greater than or equal to that in the NIR band, which can be explained by either a shock-in-jet model or the Turbulent Extreme Multi-Zone (TEMZ) model. On the other hand, five nights showed higher PD in a NIR band than an optical band, which cannot be explained by simple shock-in-jet models nor the simple TEMZ model.
The observed timescales and wavelength dependency of the PD and polarization angle suggest the existence of complicated multiple emission regions such as an irregular TEMZ model.

\end{abstract}

\section{Introduction} \label{sec:intro}

Blazars are objects characterized by intense flux variations and high polarization, and are interpreted as a relativistic jet from an active galactic nucleus (AGN) whose axis is parallel to our line of sight \citep{Urry95}.
The high luminosity and rapid variability likely stem from the relativistic Doppler boosting effects, and the high polarization between radio and optical bands is due to synchrotron radiation from the relativistic electrons accelerated in the jet.

The structure of the emission region of blazars, whether one-zone or multi-zone, is still an open question.
Past multi-wavelength studies have found correlations in the flux variations of the emission regions. 
Indeed, \cite{Marscher85} report radio and infrared flux correlations of the blazar 3C 273 and suggest a synchrotron self-Compton (SSC) model in an emission region.
They suggest a shock-in-jet scenario as a particle acceleration mechanism, assuming the blobs of relativistic plasma collide with each other and cause a first-order Fermi acceleration in the jet.
Furthermore, \cite{Aleksic12, Balokovi16, Singh22} successfully describe a number of observed blazar spectral energy distributions (SEDs) using this model.
However, some multi-wavelength studies report that the one-zone model is not suitable for reproducing the observed SEDs because it requires simple synchrotron, inverse Compton, and external Compton components \citep{MAGIC19, Yamada20, Sahu21}
, therefore a multi-zone model could be a viable alternative.

The structure of the magnetic fields, being ordered or turbulent, is also unknown \citep{Marscher85, Marscher08, Marscher14}.
\cite{Benford78} and \cite{Blandford82} discuss magnetic field models for jet formation, and conclude that the structure of the magnetic field is related to the mechanism forming the jet. As such, the structure of the magnetic field likely plays a key role in the formation of the jet.
Optical polarimetry is a reliable method to investigate the orientation and properties of the magnetic field. 
Optical radiation from blazars is linearly polarized perpendicular to the magnetic field in the emission region due to the predominance of synchrotron radiation in the optical band.
In a previous study on the optical polarization of blazars, \cite{Marscher08} reports a rotation of more than 180 degrees in the optical polarization angle in BL Lacertae (BL Lac), suggesting a magnetic field with a helical structure.
However, \cite{Marscher14} later points out that the rotation of the polarization angle can be explained by multiple emission regions with a random magnetic field.
Optical polarimetric observations on a large sample of blazars by the RoboPol program \citep{Blinov15} and Kanata telescope \citep{Abdo10,Sasada11,Ikejiri11} suggest that many blazars show intrinsic rotations in their polarization angle. 
This not only suggests fluctuations in polarization angle, but also indicates continuous variations in polarization swings.
Despite the various studies that have analyzed the emission region and magnetic field structure of blazar jets, the exact nature of these is still uncertain.

An interesting phenomena of blazar emission is the microvariability, or rapid flux variation seen within a day.
Although microvariability has been seen in several blazars \citep{Heidt96, Sasada08, Itoh18}, the exact mechanism is still not clear.
In general, the size of the emission region $R_{\rm em}$ is constrained by the variation timescale $\Delta t$ and a Doppler factor $\delta_{\rm D}$ as 
$R_{\rm em} \leq c\Delta{t}\delta_{\rm D}/(1+z)$,
where $c$ is the speed of light and $z$ is a redshift.
Since microvariability has small $\Delta t$, the resulting maximum value of $R_{\rm em}$ remains small. 
Therefore, microvariability suggests a compact emission region, which allows us to analyze fine structures that are often difficult to resolve.
Another advantage of observing microvariability is that it is free from contamination of long-term variability.
Indeed, recent high time-resolution photometric observations by {\it Kepler} and the {\it Transiting Exoplanet Survey Satellite} ({\it TESS}) have reported variations in less than a day \citep{Edelson13,Sasada17, Weaver20,Raiteri21}. 
Although there have been many studies on the microvariability of blazars, there is a scarcity of observations on their polarimetry.
Such information is useful for discussing the magnetic field in small emission regions.

Studies have shown microvariability in BL Lac \citep{Miller89,Matsumoto99,Abeysekara18}.
It has a synchrotron peak in the optical and near-infrared (NIR) bands, and thus is categorized as a low-frequency synchrotron-peaked (LSP) blazar \citep{Abdo10}.
Since August 2020, observations of BL Lac have shown its luminosity peaked in the optical and gamma-ray bands \citep{Atel_WEBT, Atel_Fermi} and showed several flares within a one-year observation period.
\cite{Kunkel21} reported the historical optical outburst where the magnitude reached $R$=11.271$\pm$0.003 mag in July 2021, much brighter than previous outburst observations in 2019 ($R$=12.80$\pm$0.01 mag, \citealp{Marchini19}) and in 2012 ($R$=13.10 mag, \citealp{Larionov12}).
In addition, {\it Fermi}-LAT recorded the highest luminosity of BL Lac in over a decade \citep{Atel_Fermi}, and optical and GeV gamma-ray flares continuously occured throughout the following year \citep{Minev21,Atel_Fermi21}.
Interestingly, BL Lac was detected in the TeV gamma-ray band (\citealt{Atel_MAGIC20,Atel_MAGIC21,Atel_LST21}), desptie it being classified as a LSP.
Put together, this flaring period is unique since the object was historically bright, flares repeatedly occurred, and TeV gamma-ray emission was detected.
Since it showed a steady degree of high polarization ($\geq$ 5$\%$) and was historically bright, observing BL Lac during an optical flaring event could provide important insight into the microvariability of blazars.

In this paper, we discuss continuous, two band observations of BL Lac during a flaring event in 2020--2021 to measure the microvariability in flux, color, and polarization.
The structure of this paper is as follows: first, we give an overview of observations and analysis methods in Section 2. In Section 3, we represent the obtained results. Lastly, we discuss the emission region and magnetic field structure of the jet in Section 4. 

\section{OBSERVATION AND ANALYSIS} \label{sec:obs_ana}

The photo-polarization data was obtained using the Hiroshima Optical and Near-InfraRed Camera (HONIR), which can perform simultaneous two-band observations \citep{Akitaya14}. 
It is installed on the 1.5-m Kanata telescope at the Higashi-Hiroshima Observatory. 
We used $R_{\rm C}$ (653.44 nm) and $J$ band (1248.5 nm) filters.
The polarimetry was performed using a fixed Wollaston prism and a rotatable half-wave plate. 
Here, the position angles of the half-wave plate in the polarimetry were 0.0, 22.5, 45.0, and 67.5 degrees.
The incident light was divided into ordinary and extraordinary light and both were recorded simultaneously, enabling us to cancel out the effect of the variable atmospheric extinction. 
The observations were carried out over 14 nights during the bright period of BL Lac, from September 2020 to August 2021.
The observation duration ranged from 1--5 hours per night and each exposure was about 30 sec.
The signal-to-noise ratio exceeded 100 in each polarization data point, enabling us to detect microvariability with an amplitude of $\sim$0.1 mag. 

We used the comparison stars BL Lac b and BL Lac f for photometric calibration \citep{comparison_Bertaud69}.
The derivations of the polarization degree (PD) and the polarization angle (PA) were the same as in \cite{Kawabata99}, where the normalized Stokes parameters $q, u$  (Stokes parameters $Q, U$ divided by $I$) were from the data of four frames at different angles of the half-wave plate.
The instrumental polarization was found to be about 0.1--0.2\%, based on the observations of the unpolarized standard star HD 14069, which is consistent with past observations with HONIR \citep{Itoh17}. 
The depolarization factor was estimated to be 0.005 ($J$ band) and 0.015 ($R_{\rm C}$ band) using a wire-grid polarizer.
The origin of the polarization angle was derived by observing strongly polarized standard stars, BD +59 389, BD +64 106, and HD 29333 \citep{Whittet92,Schmidt92}.

\section{RESULT} \label{sec:res}

\begin{figure}
    \centering
    \includegraphics[width=7cm]{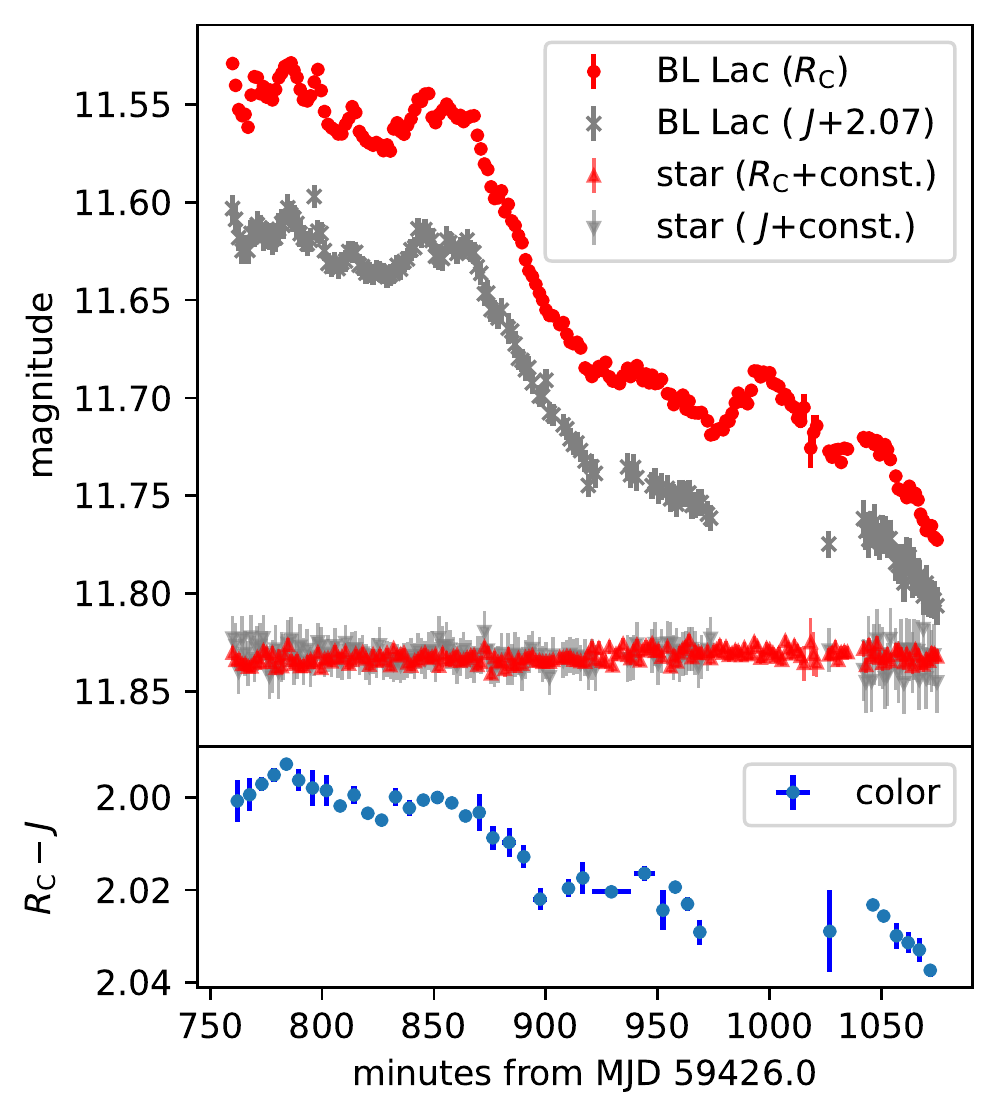}
    \caption{Time evolution of magnitude and color on July 31, 2021.
    Red dots and gray crosses in the top panel represent $R_{\rm C}$- and $J$-band magnitudes, respectively. Red triangles and gray inverted-triangles indicate the comparison star (BL Lac f).
    A part of the light curves in the $J$-band was removed due to saturation.
    Blue filled circles in the bottom panel represent the $R_{\rm C}$-$J$, calculated from the four-frame binned light curves.
    }
    \label{fig:lc}
\end{figure}

We obtained light curves from 14 observations by considering the statistical error of 0.003--0.01 mag for the photometric data. We note that most of the days showed monotonic rise or decay with amplitudes of $\sim$0.1 mag in a few hours. 
The PD varied in the range of 3.1$\pm$0.1\% to 23.7$\pm$0.2\%, and the PA varied largely overall.
In addition, mircovariability was seen on timescales of less than an hour.
Particularly, in 2021 July 31, a microvariability of the flux with a timescale of a few minutes with an amplitude 0.01--0.02 mag was found.
Figure \ref{fig:lc} shows the time variation of magnitude and $R_{\rm C}$ - $J$ color obtained on July 31, 2021. Results of the other 13 days are shown in the Appendix.
As shown in the bottom panel of Fig \ref{fig:lc}, the color variability follows the bluer-when-brighter (BWB) trend \citep{Clements01}.
In addition to other days in this work, the BWB trend was also seen in previous studies, which utilized long term \citep{Wierzcholska15} and the short term \citep{Gaur15} observations.

\begin{figure*}
    \centering
    \includegraphics[width=15cm]{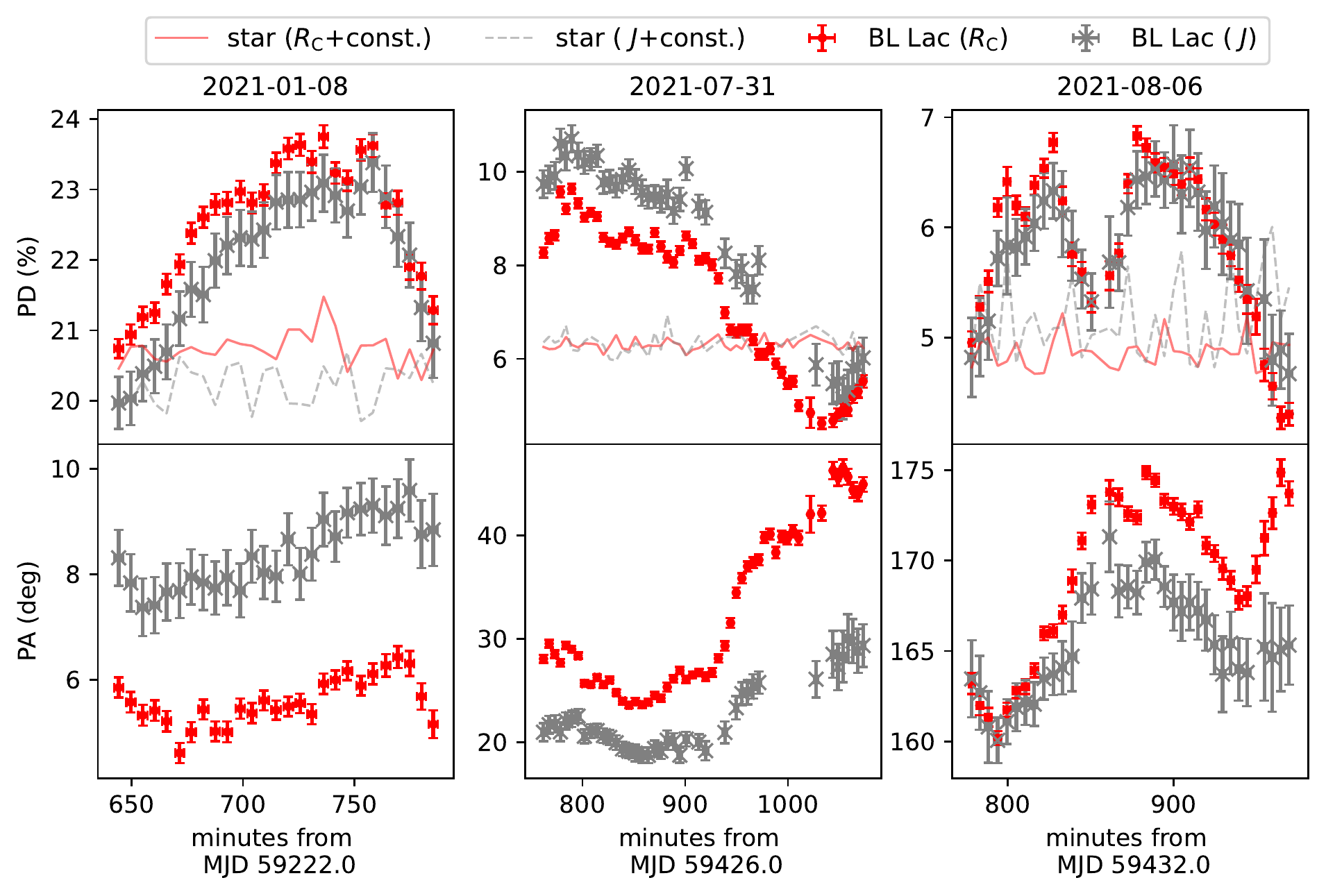}
    \caption{Time evolution of the PD and PA on January 8, 2021, July 31, 2021, and August 6, 2021.
    Red dots and gray crosses in the top panel represent $R_{\rm C}$- and $J$-band PDs, respectively.
    Red solid and gray dotted lines represent PDs of comparison star taken in the $R_{\rm C}$ and $J$ bands, respecitvely.
    The bottom panels indicate PA variations.
    }
    \label{fig:pol}
\end{figure*}

We can calculate the variability parameter $C/\Gamma$ \citep{Sasada08} to evaluate whether the variability is caused by the intrinsic variability of BL Lac or the magnitude fluctuation of the comparison star.
Here, $C$ represents the magnitude variation of the target object relative to the comparison star while $\Gamma$ is scale factor of the magnitude of the comparison star. They are defined by the following equations \citep{Romero02, Cellone07},
\begin{equation}
    C\equiv\sigma_{\rm bl}/\sigma
\end{equation}
and\\
\begin{equation}
\Gamma^2\equiv \left(\frac{I_1}{I_{\rm bl}}\right)^2 \frac{I_1^2(I_{\rm bl}+p)+I_{\rm bl}^2(I_1+p)}{I_2^2(I_{\rm bl}+p)+I_{\rm bl}^2(I_2+p)},\\
 \\
 \\
 \\
\end{equation}
where the parameter $\sigma_{\rm bl}$ is the standard deviation of ${\log}_{10}I_{\rm bl}-{\log}_{10}I_2$ and $\sigma$ is the standard deviation of ${\log}_{10}I_1-{\log}_{10}I_2$.
The $I_{\rm bl}$ is the counts of BL Lac, while $I_1$ and $I_2$ are the counts of the two comparison stars, respectively.
The term $p$ is equal to $n(s+r^2)$, where $n$ is the total number of pixels in the aperture, $s$ is the median of the sky counts, and $r$ is the readout noise. 
A variability parameter of $C/\Gamma = 2.576$ indicates a significant variability with a 99\% confidence level \citep{Romero02}, and thus we set the threshold for distinguishing the microvariability to $C/\Gamma = 2.576$.
The variation parameter $C/\Gamma$ shows various values depending on the source, and can range from $<$1 to $>$30 \citep{Cellone07}.
Based on this indicator, significant variations were detected in ten of the fourteen data sets.
Table \ref{tab:table1} lists $C/\Gamma$, the mean magnitude in each filter, and the spectral index of each observation.
\begin{longrotatetable}
\begin{threeparttable}
    \caption{The significance of variability and spectral indices.}
    \begin{tabular}{c c c c c c c c}
    \hline \hline
    Obs. Date & MJD Start--MJD Stop &
        $C/\Gamma$\tnote{a} & $R_{\rm C}$\tnote{b} & $J$\tnote{c} & $\alpha$\tnote{d} &  Bands\tnote{e} & Timescales\tnote{f} \\
        (yy-mm-dd) &&&[mag]&[mag]&&&[minutes] \\
        \hline 
       2020-09-21 & 59113.73259--59113.79096 & 0.81 (N) & 12.633 (0.003) & 10.656 (0.008) & 1.126 $\pm$ 0.004 & $R_{\rm C}$ & --\\
        2020-09-22 & 59114.53215--59114.64102 & 6.38 (V) & 12.382 (0.023) & 10.426 (0.021) & 1.097 $\pm$ 0.005 & $R_{\rm C}$ & 19 $\pm$ 8\\
        2020-10-02 & 59124.44197--59124.59158 & 11.48 (V) & 12.291 (0.057) & 10.336 (0.041) & 1.095 $\pm$ 0.005 & $R_{\rm C}$ & 37 $\pm$ 6\\
        2020-10-12 & 59134.46969--59134.70379 & 1.22 (N) & 12.529 (0.011) & 10.560 (0.011) & 1.115 $\pm$ 0.004  & $R_{\rm C}$ & --\\
        2021-01-08 & 59222.44739--59222.54541 & 2.24 (N) & 11.899 (0.038) & 9.875 (0.031) & 1.194 $\pm$ 0.007 & $R_{\rm C}$ & --\\
        2021-01-20 & 59234.42020--59234.50068 & 2.39 (N) & 12.157 (0.021) & 10.096 (0.017) & 1.245 $\pm$ 0.006 & $J$ & --\\
        2021-02-03 & 59248.42281--59248.47106 & 6.96 (V) & 11.780 (0.041) & 9.774 (0.034) & 1.167 $\pm$ 0.009 & $R_{\rm C}$ & 20 $\pm$ 2\\
        2021-07-25 & 59420.68825--59420.75068 & 6.72 (V) & 12.148 (0.022) & 10.034 (0.018) & 1.322 $\pm$ 0.006 & $J$ & 31 $\pm$ 6\\
        2021-07-26 & 59421.57541--59421.66953 & 14.17 (V) & 12.070 (0.025) & 9.997 (0.022) & 1.263 $\pm$ 0.006 & $R_{\rm C}$ & 16 $\pm$ 2\\
        2021-07-27 & 59422.62880--59422.70254 & 7.83 (V) & 11.812 (0.018) & 9.784 (0.012) & 1.200 $\pm$ 0.008 & $R_{\rm C}$ & 19 $\pm$ 2\\
        2021-07-31 & 59426.52923--59426.74524 & 46.42 (V) & 11.643 (0.076) & 9.617 (0.066) & 1.198 $\pm$ 0.009 & $J$ & 5 $\pm$ 1\\
        2021-08-03 & 59429.59383--59429.80045 & 24.42 (V) & 11.454 (0.040) & 9.489 (0.033) & 1.110 $\pm$ 0.011 & $J$ & 10 $\pm$ 2\\
        2021-08-06 & 59432.54048--59432.67341 & 16.09 (V) & 11.175 (0.017) & 9.215 (0.014) & 1.103 $\pm$ 0.014 & equal & 14 $\pm$ 2\\
        2021-08-09 & 59435.48704--59435.67907 & 10.85 (V) & 11.509 (0.015) & 9.482 (0.011) & 1.198 $\pm$ 0.010 & $J$ & 21 $\pm$ 5\\
        \hline
    \end{tabular}
    \begin{tablenotes}
    \item[a] Variability parameters which were obtained from $R_{\rm C}$ band data. The (V) indicates variable and (N) indicates non-variable.
    \item[b,c] The means of magnitude taken by both bands and the numbers in parentheses represent the standard deviations.
    The magnitudes are reported in the Vega system.
    \item[d] Spectral index $\alpha$. It is defined as $f$ $\propto \nu^{-\alpha}$, where $f$ is the flux [jy] and $\nu$ is the frequency [Hz]. The Galactic extinction corrections for the $R_{\rm C}$- and $J$- bands are 0.719$\pm$0.030 and 0.247$\pm$0.010 mag, respectively \citep{Cardelli89}.
    \item[e] The bands which showed higher PD.
    \item[f] Timescales estimated by equation (3).
    \label{tab:table1}
    \end{tablenotes}
    \end{threeparttable}
\end{longrotatetable}
We also found differences in the PD and PA when comparing their $R_{\rm C}$ and $J$ band values; nine nights showed higher PD in the optical band than the NIR band, while five nights showed higher PD in the NIR band than in the optical band.
Figure \ref{fig:pol} shows the time variation of the PD and PA on January 8, 2021, July 31, 2021, and August 6, 2021.
For example, on 2021 January 8, we found higher PD values in the $R_{\rm C}$ band than in the $J$ band. 
This is similarly seen on seven other days, where the mean difference in PD is 0.91\%. 
On the other hand, on July 31, 2021 and four other days, we found lower PD values in the $R_{\rm C}$ band than the $J$ band by a mean difference of 0.75\%.
Only for observation on August 6, 2021 did PDs from both bands match; the PAs, however, were not equivalent.
We can see variations of PD and PA with timescales of several hours.
The difference in PA between the two bands was also observed, but no relation with the PD of the comparison star is seen.
Therefore, this difference in PD and PA between the $R_{\rm C}$ and $J$ bands is intrinsic rather than a systematic error.

\begin{figure*}
    \centering
    \includegraphics[width=14cm]{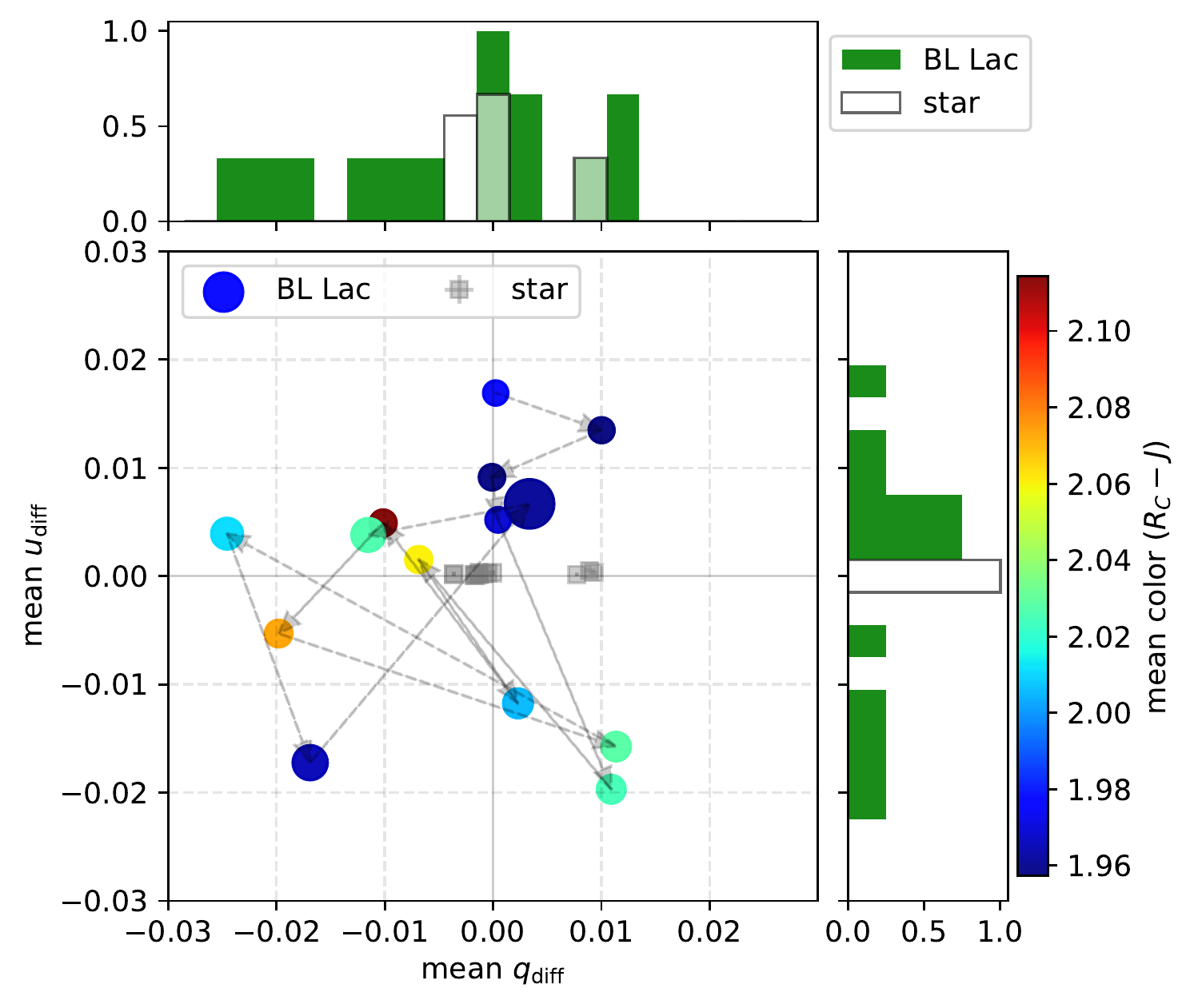}
    \caption{
    Scatter plot of differential $q$ and $u$ values between $R_{\rm C}$ and $J$ ($q_{\rm diff}$ and $u_{\rm diff}$) for each observation. Each data point shows the average value in each observation.
    Gray squares represent data of the comparison star and filled circles represent data of BL Lac.
    The circle size represents the flux value for each data point.
    The colors of the circles represent the $R_{\rm C}-J$ color, where values are listed in the color bar. 
    Gray dashed arrows that connected each data point indicate time order.
    Normalized histograms on the horizontal and vertical axes represent projections along the $q_{\rm diff}$ and $u_{\rm diff}$ axis, respectively.
    }
    \label{fig:hist}
\end{figure*}

Next, we calculated the difference in the normalized Stokes parameters between the $R_{\rm C}$ and $J$ bands.
We define $q_{\rm diff}$ and $u_{\rm diff}$ as the difference of $q$ and $u$ values between the $R_{\rm C}$ and $J$ bands (i.e. $q_{\rm diff}$ = $q_{\rm R_{\rm C}}-q_{\rm J}$, where $q_{\rm R_{\rm C}}$ and $q_{\rm J}$ are $q$ in $R_{\rm C}$ and $J$ band, respectively).
Figure \ref{fig:hist} shows a distribution of average $q_{\rm diff}$ and $u_{\rm diff}$ calculated for BL Lac and the comparison stars.
The $u_{\rm diff}$ of the comparison star is concentrated around 0.0, while that of BL Lac is spread from -0.02 to 0.02.
This means that there was a significant wavelength dependence of the PD and PA for BL Lac.
We find no clear dependence of $q_{\rm diff}$ and $u_{\rm diff}$ on the flux and the $R_{\rm C}-J$ color.

\section{DISCUSSION} \label{sec:dis}

We detected variations on timescales of roughly a few hours in ten of the fourteen observations, where one of them showed variations with short timescales on the order of minutes.
The timescale $\Delta t$ can be estimated from the following equation \citep{Romero02}:
\begin{equation}
    \Delta t = \frac{1}{1+z}\frac{\Delta F}{dF/dt},
\end{equation}
where $z$ is the redshift ($z = 0.069$, \citealp{Miller77}) and $\Delta F$ is the peak of the flux in which the baseline component is subtracted ($\Delta F = 4.60 \times 10^{-12}$ erg cm$^{-2}$ s$^{-1}$) as follows.
As shown in Figure \label{fig:estimate_deltat} top, we defined the baseline flux history using the 3rd--6th time bins as well as those after the 13th time bin, as represented by the line.
Then we calculated the $\Delta F$ (“peak flux” minus “baseline flux”) at the time when the flux reaches its peak value.
For $dF/dt$, we defined the rising and decaying period by two lines as shown in Figure 4 top (from the 7th to the 13th time bin) and obtained their slopes as $dF/dt$.
\begin{figure}
    \centering
    \includegraphics[width=7cm]{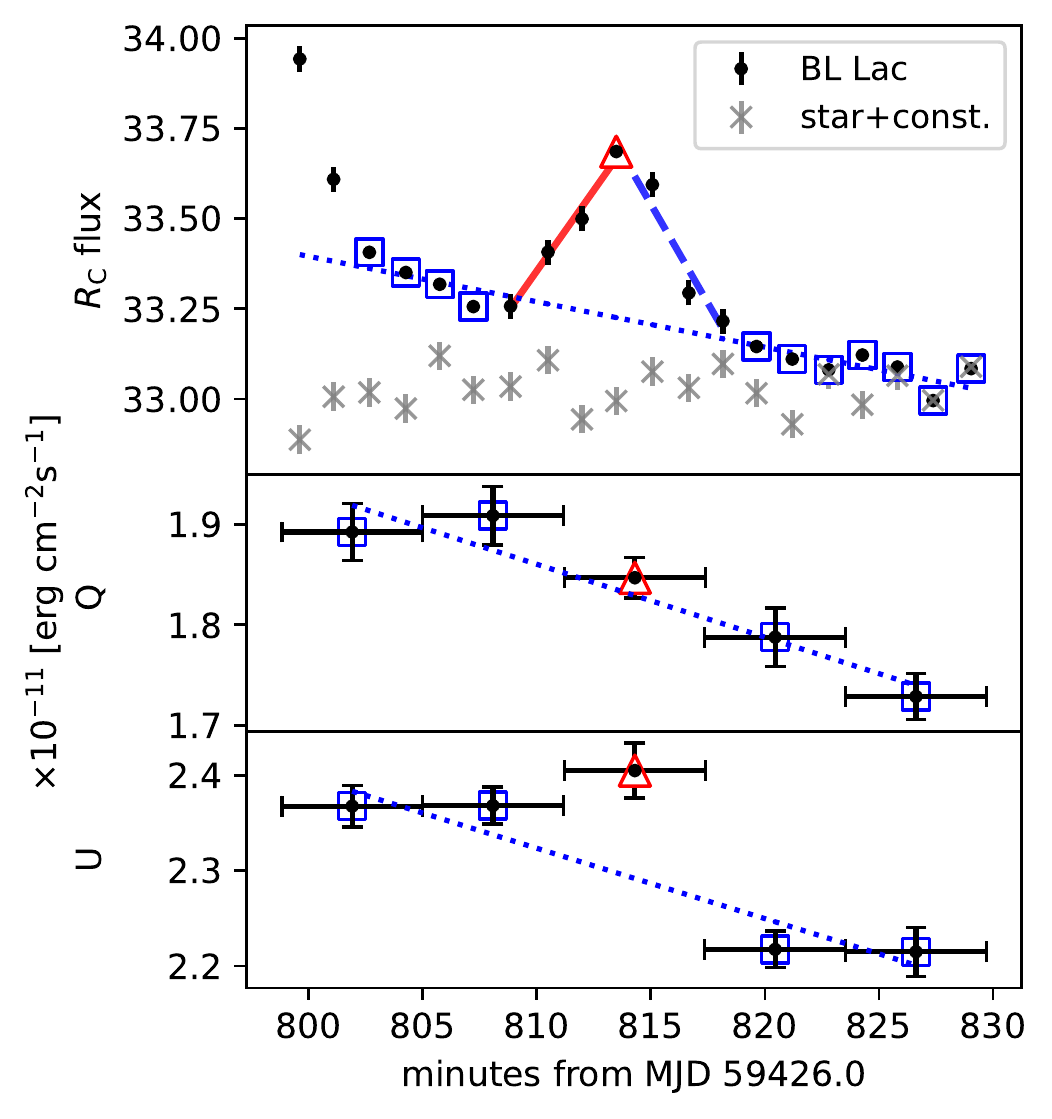}
    \caption{Top panel: the microvariability of the flux detected in $R_{\rm C}$ band on July 31, 2021.
   Middle and Bottom panels: the time variation of the Stokes parameter $Q$ and $U$, respectively. 
    Each of the $Q$ and $U$ bins corresponds to four frame sets (a cycle of the half wave plate) of the flux lightcurve. 
    Blue dotted lines represent the baseline.
    Red triangles indicate peak of the microvariability.
    Red solid line and blue dashed line in the top panel represent the linear approximation for the rising and decaying periods, respectively.
    }
    \label{fig:estimate_deltat}
\end{figure}
Based on this equation, the rise time is determined to be 340 $\pm$ 45 seconds and the decay time to be 278 $\pm$  seconds for 2021 July 31. Note that we considered only the statistical errors.
We subtracted the baseline and obtained $C/\Gamma$=5.21 for this variability, and thus found this variation to be instrinsic.

The maximum size of the emission region $R_{\rm em}$ can be estimated from the light crossing time,
$R_{\rm em} \leq c\Delta{t}\delta_{\rm D}/(1+z)$.
From equation 3, $\Delta{t}$ for both rise and decay phase are about 5 minutes.
The Doppler factor $\delta_{\rm D}$ is estimated to be $7.2 \pm 1.1$ by the kinematics study using Very Long Baseline 
Interferometry (VLBI) \citep{Jorstad05}, and thus the maximum $R_{\rm em}$ is constrained to be $(0.6\pm0.2)\times10^{12}$ m.
\cite{Capetti10} estimated a central black hole mass of BL Lac using three different methods; correlation between black hole mass and NIR luminosity of the host galaxy \citep{Marconi03}, reverberation-mapping technique \citep{Bentz09}, and a spectroscopic study \citep{Peterson04,Kaspi05,Vermeulen95}. 
From these studies, $M_{\rm BH}$ is estimated to be $6\times10^8M_{\odot}$, $3.76^{+1.28}_{-0.95}\times10^8M_{\odot}$, and $2\times10^7M_{\odot}$, respectively.
These give the Schwarzschild radius $R_{\rm s}$, from which $R_{\rm em}$ becomes $(0.3\pm0.1) R_{\rm s}$, $(0.55\pm0.23) R_{\rm s}$, and $(10\pm3) R_{\rm s}$, respectively.
The core size in the BL Lac jet was estimated to be (1.15--1.76)$\times10^{15}$ m (which equates to (6.5--9.9)$\times10^{2}R_{\rm s}$ with $M_{\rm BH}=6\times10^8M_{\odot}$) based on 43--86 GHz observations \citep{Casadio21}, and thus the estimated emission region is smaller than the radio core. 
\cite{Bottcher13} performed SED fitting using a leptonic one-zone SSC and external Compton model, and they obtained a minimum variability timescale to be 1.8 hours (i.e, $R_{\rm em}$=13.1$\times10^{12}$ m). Therefore, the one-zone model cannot explain the five-minute microvariability.

Past studies suggested that the changing viewing angle can explain the blazar flux and PD variations \citep{Hagen-Thorn08}. Let us discuss if this model can explain our results.
When the viewing angle changes, the change of photon energy is proportional to the doppler factor $\delta_{\rm D}$, whereas the change in flux approximately depends on $\delta_{\rm D}^3$ (see equation 1 in \citealp{Hagen-Thorn08}).
The variation of color should be negligible relative to the variations in flux, making it difficult to explain our results.

\begin{figure}
    \centering
    \includegraphics[width=7cm]{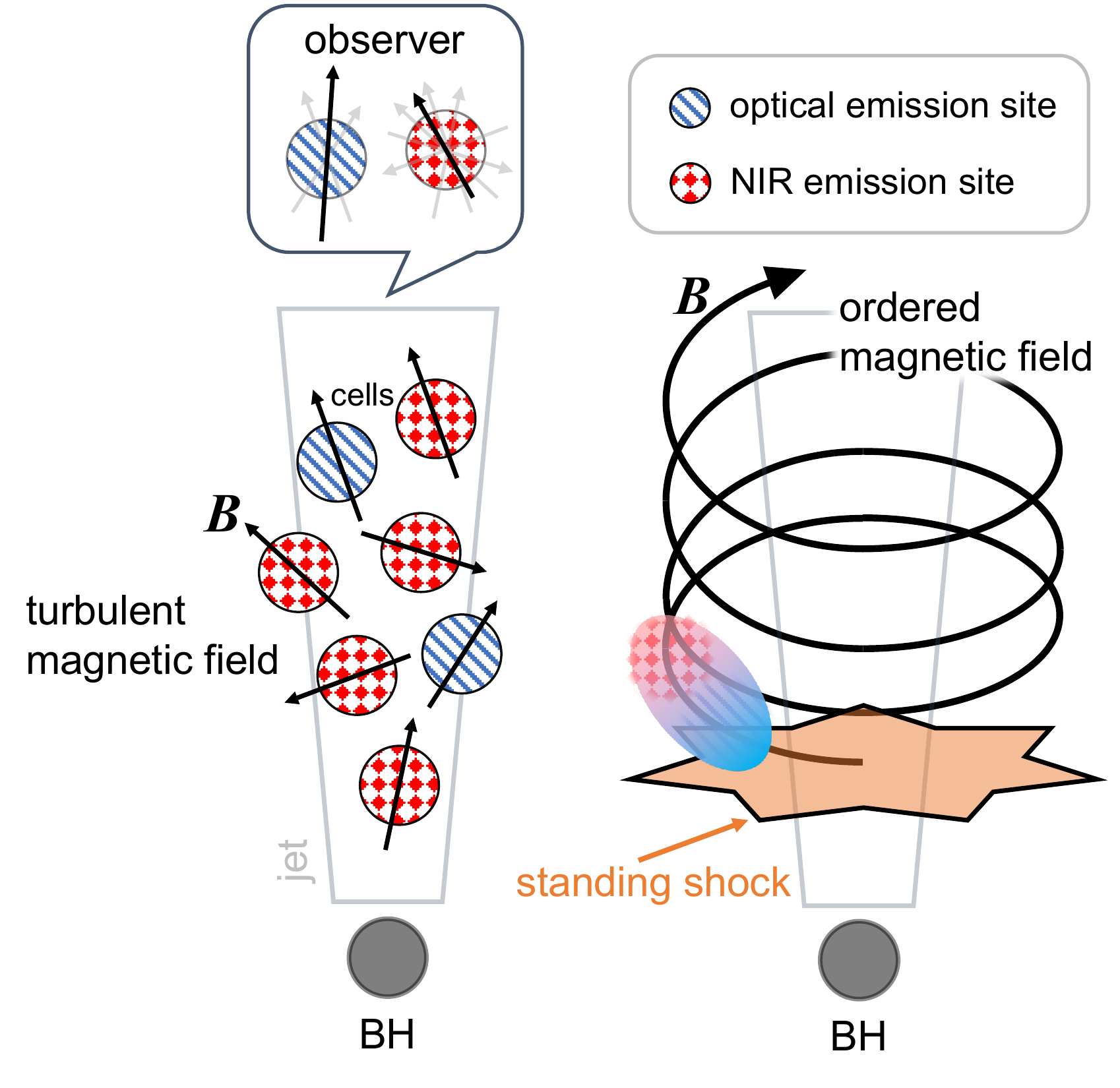}
    \caption{Schematic illustration summarizing the discussion in Section 4. 
    On the left is a diagram of the differential PD due to turbulent magnetic fields and multi-zone radiation. 
    On the right is a diagram of the differential PA due to the ordered (i.e., helical) magnetic field. 
    The checked blue and hatched red regions in the figure represent the dominant emission regions in optical and NIR, respectively. 
    Solid arrows indicate the direction of the magnetic field.}
    \label{fig:sketch}
\end{figure}

We conclude that the shock-in-jet model does not explain the observed microvariability, we next discuss whether the differential values of $Q$ and $U$ in the optical and NIR bands can be explained by this model.
The PD is larger in the optical than in the NIR on some days, and vice versa on others.
In the shock region, the magnetic field is stronger than in the unshock region due to shock compression. 
In addition, the shock-accelerated maximum electron energy is higher than the unshock one.
Thus, as the emission region moves downstream of the jet, the magnetic field should become weaker and less ordered by free expansion, as shown in Figure \ref{fig:sketch} (right).
As a result, both the emitted photon energy and PD should decrease \citep{Angelakis16}.
High energy electrons emit photons at higher frequencies, and thus the shock-in-jet scenario can explain the higher PD in the optical band.
Since the emission region is different between the optical and NIR, the PA could be different between two wavelength regimes.
However, in the case of higher PD in the NIR band than the optical cannot be explained by this model. 

Next, we discuss about the Turbulent Extreme Multi-Zone (TEMZ) model
\citep{Marscher14,Marscher17,Marscher21} which explains the microvariability and wavelength-dependence of the polarization. 
The TEMZ model assumes separated multiple emission regions or `cells' in the jet (Figure \ref{fig:sketch}, left).
In each cell, the energy power spectrum of the electrons and the direction of the magentic field are uniform.
However, the energy power spectrum of the electrons is different across cells, and thus the emitting photon 
spectrum also becomes different from cell to cell.
Polarized emissions from the cells cancel each other out because of the turbulent magnetic fields and the complicated nature of the $Q$ and $U$ components.
Therefore, the PD from the cells is decreased, depending on the number of cells.
As shown in Table \ref{tab:table1}, spectral index $\alpha$ indicates a steep spectrum ($\alpha > 1$) and thus the number of high-energy electrons should be small.
The average value of the spectral index overall observations is $\alpha$ = 1.17, leading to a photon index $\Gamma$ = $\alpha$+1 = 2.17.
Considering that the ratio of the frequency of $R_{\rm C}$ and $J$ is 1.9, NIR photons are about 4 times more numerous than optical photons. 
Therefore, we can assume that most cells emit NIR photons while only a handful of cells emit optical photons.
In that case, the PD in the NIR band should approach zero due to a large number of NIR-emitting cells, while PD in the optical band should not be canceled out as much due to a small number of optical-emitting cells. 
The difference in PA between the two bands can also be explained by this model \citep{Marscher21}.
Therefore, the TEMZ model can better explain the high PD in the optical than in the NIR.

Nine of the fourteen observations showed that the PD in the optical band is higher than or comparable to that in the NIR (e.g. 2021 January 8). This is consistent with the above explanation.
However, we found five days that showed higher PD in the NIR than the optical, and this trend cannot be explained by the above scenario alone.
For example, we could consider the scenario where the turbulent magnetic field may have been ordered accidentally at a bright spots (cells) in the NIR.
The turbulent plasma has different magnetic field directions, and thus a wavelength dependence of PA could be caused.
Another scenario is that one of the cells which is bright in the NIR and accidentally becomes much brighter than others.
We could also explain the higher PD in the NIR band by a two-zone shock-in-jet model, where the emission from the zone with higher PD is brighter in the NIR while the lower PD is in the optically bright zone.

In the following paragraphs, we will discuss the polarization behavior in the microvariability on July 31, 2021.
We defined ($Q_{\rm base}(t)$,$U_{\rm base}(t)$) as shown as the blue dotted lines in Figure 4, and obtained the microvariability polarization component ($Q_{\rm var}(t)$,$U_{\rm var}(t)$) as follows,
\begin{equation}
    Q_{\rm obs}(t) = Q_{\rm base}(t) + Q_{\rm var}(t),\\
    U_{\rm obs}(t) = U_{\rm base}(t) + U_{\rm var}(t),
\end{equation}
where ($Q_{\rm obs}(t)$,$U_{\rm obs}(t)$) are the observed polarization component.
We defined the polarized flux $PF(t)$ which is PD calculated from $\sqrt{Q_{\rm var}^2(t)+U_{\rm var}^2(t)}$, and the differential flux $\Delta F$ is obtained from equation 3.
Next, we calculated the PD in peak of the microvariability component: $p_{\rm peak}$, which can be calculated from the following equation,
\begin{equation}
     p_{\rm peak} = \frac{PF_{\rm peak}}{\Delta F_{\rm peak}},
\end{equation}
where $\Delta F_{\rm peak}$ is a peak of $\Delta F$ and $PF_{\rm peak}$ is a polarized flux at that time.
In this case, we chose the peak time to be from the 9th to 12th bin in the flux lightcurve.
As a result, we obtained $p_{\rm peak}$ of $R_{\rm C}$ to be $38\pm11 \%$ for 2021 July 31.
For comparison, we calculated the mean of observed PD during the period where only the baseline component is obsereved, which resulted in $8.81\pm0.08\%$, where the uncertainty represents the standard error.
The $p_{\rm peak}$ is higher than the mean PD, suggesting that the microvariability comes from a region with an ordered magnetic field.
For the BL Lac-type object S5 0716+714, \cite{Sasada08} calculated the PD of the microvariability component to be 27$\pm$5 \%, consistent with our results.
Our measured microvariability shows a high PD, which suggests emission from inside the cell.

In summary, both the simple shock-in-jet model and the TEMZ model can explain the higher PD found in the optical band than in the NIR, and the difference in PA between the two bands.
The microvariability observed cannot be explained by the shock-in-jet model alone, but can be explained by the TEMZ model.
As such, we conclude that the TEMZ model is suitable to explain our overall results.
However, the higher PD seen in the NIR than in the optical cannot be explained either by an ordered magnetic field model, or the TEMZ model. A two-zone shock-in-jet model or a more complicated variation of the TEMZ model is needed to explain these results.

\section{CONCLUSION} \label{sec:con}
In this study, several simultaneous optical and NIR continuum observations were performed for BL Lac during an outburst period between 2020 and 2021.
The light curves suggest compact emission regions and microvariability accompanied by particle acceleration.
While the observed microvariability cannot be explained by the one-zone shock-in-ject model, it can be explained by the TEMZ model.
Observed PDs and fluxes of BL Lac do not clearly correlate with each other, and the PD and PA of the optical and NIR bands differ except for one observation.
We measure higher PD in the optical band than in the NIR, which can be explained by both models.
However, in other observations, we measure higher PD in the NIR band, which cannot be explained by either of these simple models, suggesting a more complicated situation such as a two-zone shock-in-jet model or an irregular TEMZ mode.

\noindent
\begin{acknowledgements} \label{sec:acknowledgement}
{\large Acknowledgements} \\
We would like to thank Alan P. Marscher and Svetlana Jorstad for their helpful advice.
This work was supported by Japan Society for the Promotion of Science (JSPS) KAKENHI Grant Numbers JP19K14761, and JP21H01137.
This work was also partially supported by Optical and Near-Infrared Astronomy Inter-University Cooperation Program from the Ministry of Education, Culture, Sports, Science and Technology (MEXT) of Japan.
We are grateful to the observation and operating members of Kanata Telescope.
R.I. is supported by JST, the establishment of university fellowships towards the creation of science technology innovation, Grant Number JPMJFS2129.
This research was supported by a grant from the Hayakawa Satio Fund 
awarded by the Astronomical Society of Japan.
\end{acknowledgements}

\def\aap{A\&A}

\bibliography{temp_body}{}
\bibliographystyle{aasjournal}
\newpage
\appendix
\section*{Other light curves and variation of $q$ and $u$}

Figure \ref{fig:lc_appendix1} shows light curves not shown in the main text. We used a linear fit of flux versus time for the timescale estimation (Table \ref{tab:table1}).
In Figure \ref{fig:pol_appendix1}, we show the variations in the PA and PD.
We can see the significant variation in PA and PD, and we calculated the standard deviation of the $q$ and $u$ parameters for BL Lac and the comparison star BL Lac f.
To estimate variability for $q$ and $u$ quantitatively, we calculated $C\equiv\sigma_{\rm bl}/\sigma$, where $\sigma_{\rm bl}$ is the standard deviation of $q$ or $u$ for BL Lac, and $\sigma$ is for the comparison star (BL Lac f), respectively. We defined to variable as $C$ exceeds 2.576 or not.
We list the values of $C$ for $q$ and $u$ in Table \ref{tab:qu}.
We note that four dates showed variability in both $q$ and $u$.

\begin{figure}
    \centering
    \includegraphics[width=16cm]{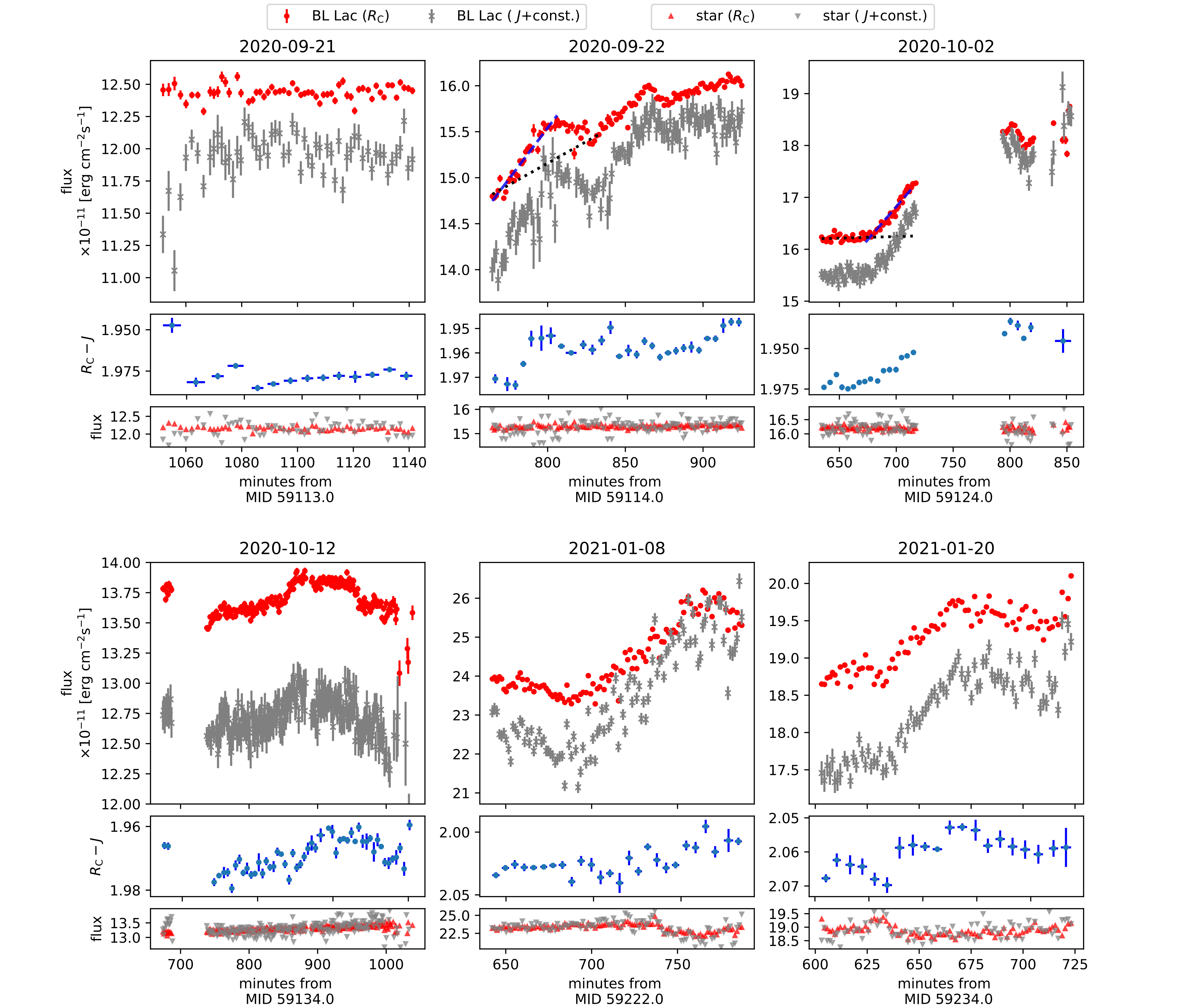}
    \caption{The top panels show BL Lac light curves for each band.Black dotted- and blue dashed- lines indicate baseline and variable components using linear fitting of flux versus time for $R_{\rm C}$ band data, respectively.The 2nd panels show color ($R_{\rm C}-J$) evolution and the bottom panels show time evolution in flux for the comparison star (BL Lac f), respectively.}
    \label{fig:lc_appendix1}
\end{figure}

\begin{figure}
    \centering
    \includegraphics[width=16cm]{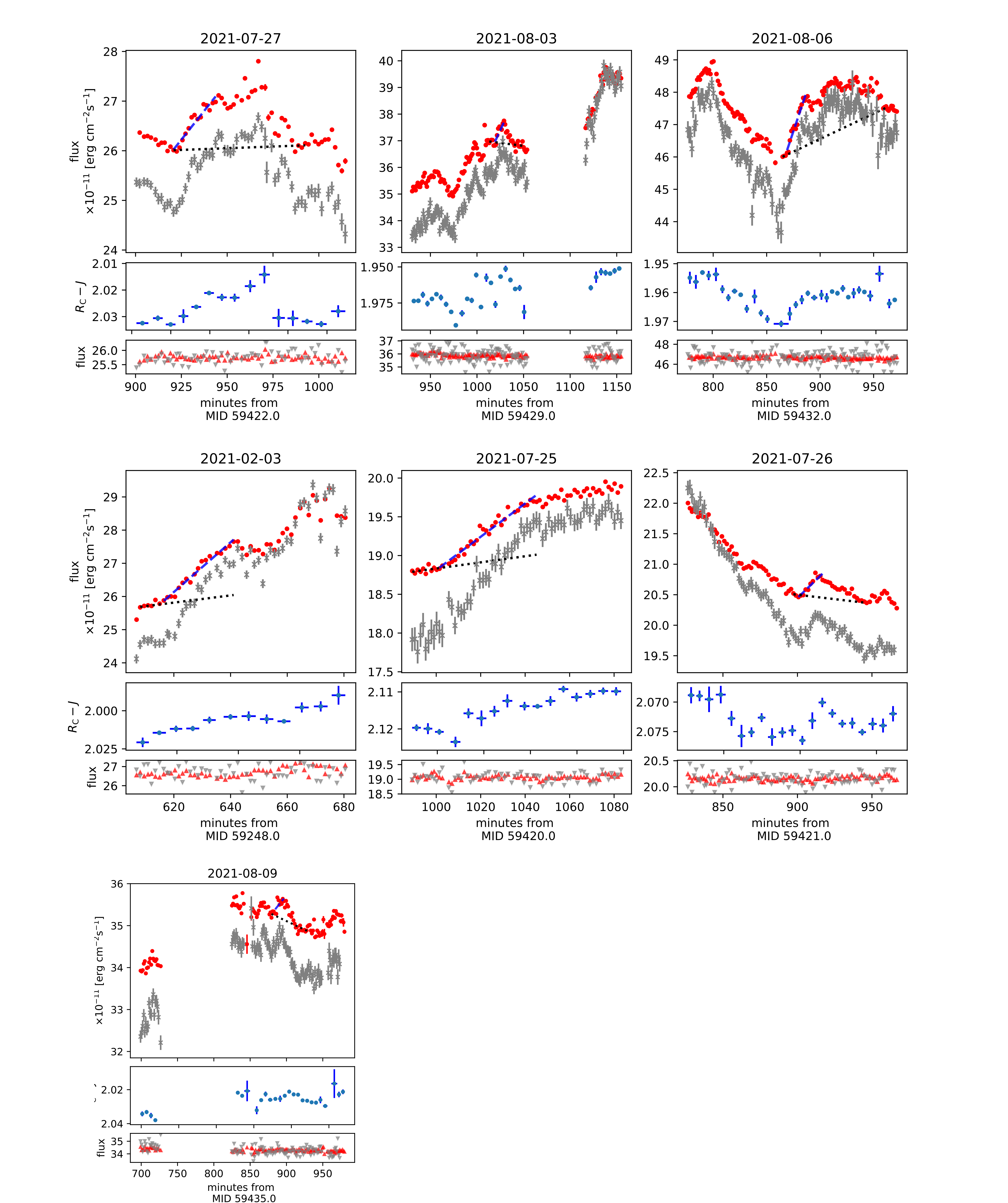}
    \caption{(Cont.).}
    \label{fig:lc_appendix1}
\end{figure}

\begin{table}[t]
\begin{threeparttable}
    \caption{Variability of the $q$ and $u$ in $R_{\rm C}$ band}
    \begin{tabular}{c c c c}
    \hline \hline 
        Obs. Date & $C_q$\tnote{a} &
        $C_u$\tnote{b} & Variable or not\tnote{c}\\
        (yy-mm-dd) &&& ($C_q$,$C_u$) \\
        \hline
        2020-09-21 & 1.292 & 1.158 & (N,N) \\
        2020-09-22 & 2.223 & 5.296 & (N,V) \\
        2020-10-02 & 4.097 & 5.432 & (V,V) \\
        2020-10-12 & 0.642 & 4.911 & (N,V) \\
        2021-01-08 & 3.482 & 1.633 & (V,N) \\
        2021-01-20 & 2.013 & 4.663 & (N,V) \\
        2021-02-03 & 1.854 & 4.623 & (N,V) \\
        2021-07-25 & 3.218 & 2.441 & (V,N) \\
        2021-07-26 & 7.106 & 2.286 & (V,N) \\
        2021-07-27 & 4.413 & 1.287 & (V,N) \\
        2021-07-31 & 19.808 & 6.935 & (V,V) \\
        2021-08-03 & 5.451 & 6.808 & (V,V) \\
        2021-08-06 & 3.794 & 7.106 & (V,V) \\
        2021-08-09 & 0.860 & 4.849 & (N,V) \\
        \hline
    \end{tabular}
       \begin{tablenotes}
    \item[a,b] Variability parameter obtained from $R_{\rm C}$ band data. 
    \item[c] Exceed 2.576 (V) or not (N) for $q$ and $u$, respectively.
    \label{tab:qu}
    \end{tablenotes}
    \end{threeparttable}
\end{table}

\begin{figure}
    \centering
    \includegraphics[width=14cm]{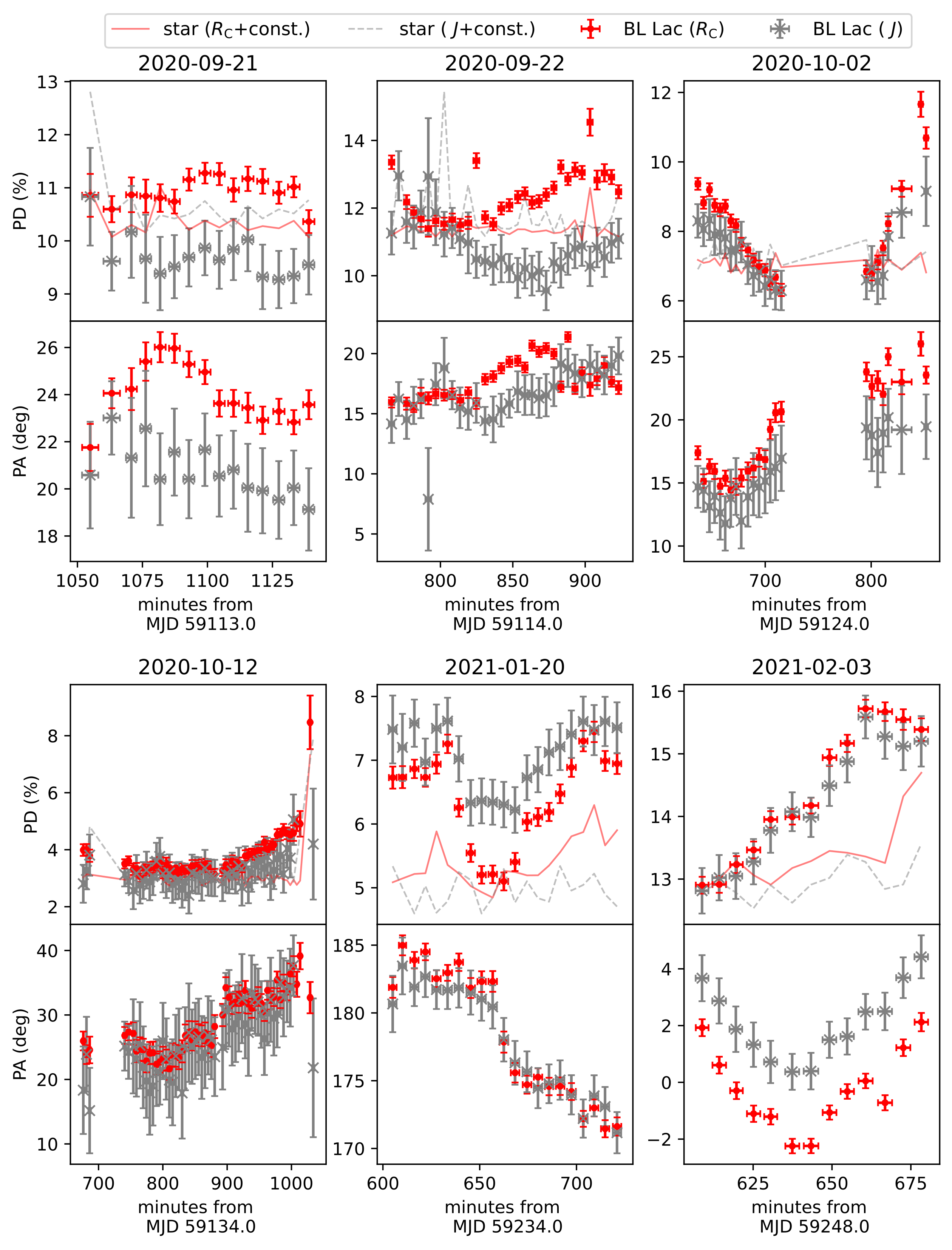}
    \caption{Time evolution of PD and PA for each observation. Red points and gray crosses indicate BL Lac data taken by $R_{\rm C}$- and $J$-band, and solid and dashed-lines for the comparison star (BL Lac f).}
    \label{fig:pol_appendix1}
\end{figure}

\begin{figure}
\addtocounter{figure}{-1}
    \centering
    \includegraphics[width=13cm]{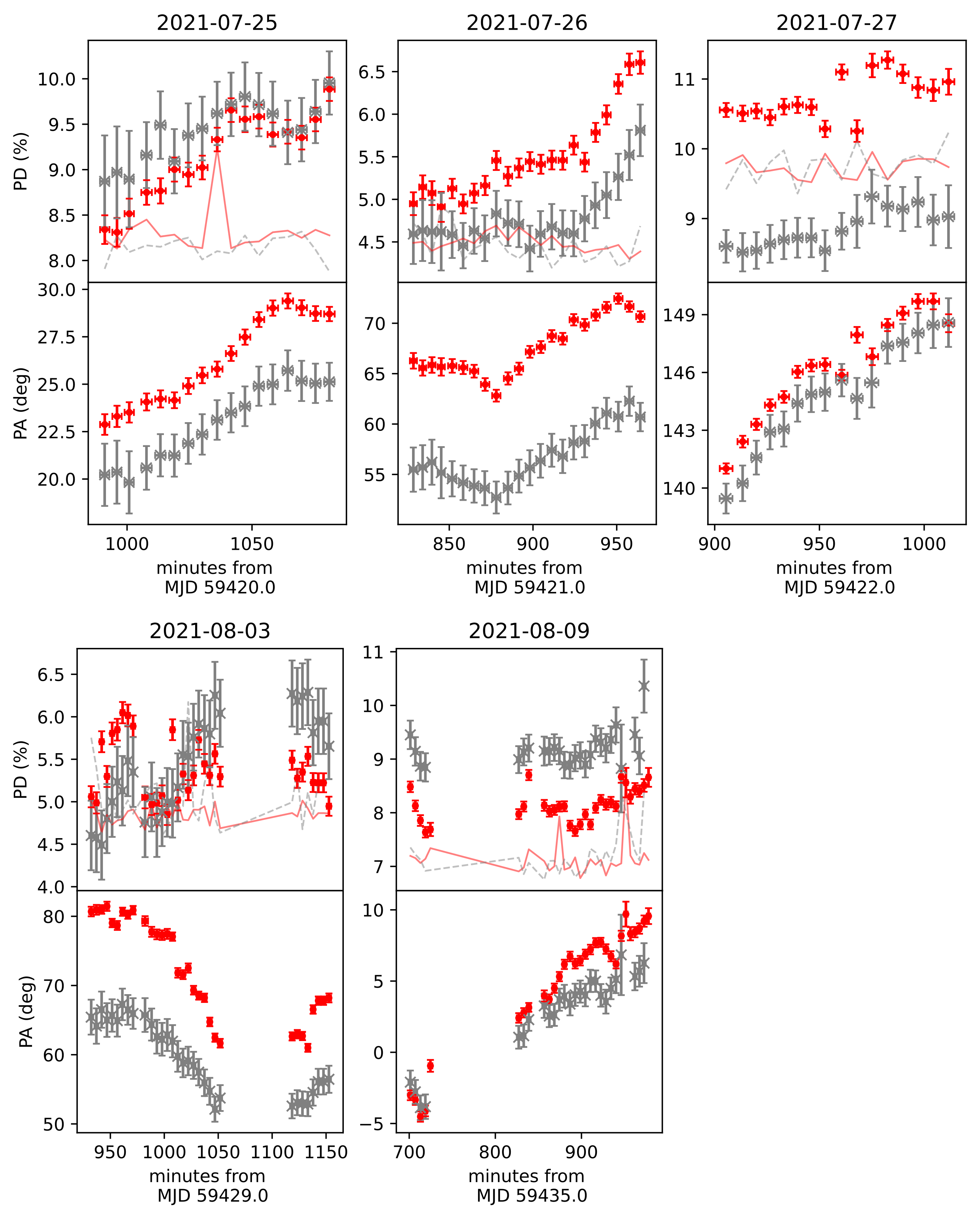}
    \caption{(Cont.).}
    \label{fig:pol_appendix2}
\end{figure}

\end{document}